\documentclass[submitted,5p,times,twocolumn]{elsarticle}

\usepackage{subfigure,dcolumn}
\usepackage{epstopdf}
\usepackage{mhchem}
\usepackage{upgreek}
\usepackage{graphicx}
\usepackage{subfigure}
\usepackage{multirow}
\usepackage{footnote}
\usepackage{setspace}
\usepackage{ulem} 
\usepackage[dvipsnames]{xcolor}
\usepackage{physics}
\usepackage{xparse}
\usepackage{amsmath}
\usepackage{cuted}
\usepackage{color}
\usepackage{float}

\usepackage{float}
\usepackage{stfloats}
\usepackage{graphicx}
\usepackage{booktabs} 
\usepackage{setspace}
\usepackage{tabularx}
\usepackage[colorlinks,
            linkcolor=blue,
            anchorcolor=blue,
            citecolor=blue]{hyperref}

\usepackage{caption} 
\def \esym{$E_{\rm sym}(\rho)$}

\begin{document}
\begin{frontmatter}

\title{An FPGA-based Trigger System for CSHINE}

\author[THU]{Dong Guo\corref{cor1}} 
\author[THU]{Yuhao Qin}   
\author[THU]{Sheng Xiao}  
\author[THU]{Zhi Qin}     
\author[THU]{Yijie Wang}  
\author[THU]{Fenhai Guan} 
\author[THU]{Xinyue Diao} 
\author[THU]{Boyuan Zhang} 
\author[THU]{Yaopeng Zhang} 
\author[THU]{Dawei Si}    
\author[IMP]{Shiwei Xu} 
\author[IMP]{Xianglun Wei} 
\author[IMP]{Herun Yang} 
\author[IMP]{Peng Ma} 
\author[IMP]{Tianli Qiu} 
\author[IMP,LZU]{Haichuan Zou} 
\author[IMP]{Limin Duan} 
\author[THU]{Zhigang Xiao\corref{cor2}}
\address[THU]{Department of Physics, Tsinghua University, Beijing 100084, China}
\address[IMP]{Institute of Modern Physics, Chinese Academy of Science, Lanzhou 730000, China}
\address[LZU]{School of Nuclear Science and Technology, Lanzhou University, Lanzhou 730000, China\vspace{-3.5em}}
\cortext[cor1]{guodong19@mails.tsinghua.edu.cn (Dong Guo)}
\cortext[cor2]{xiaozg@tsinghua.edu.cn (Zhigang Xiao)}

\begin{abstract}
A trigger system of general function is designed  using the  commercial module CAEN V2495 for heavy ion nuclear reaction experiment at Fermi energies. The system has been applied and verified on CSHINE (Compact Spectrometer for Heavy IoN Experiment). Based on the field programmable logic gate array (FPGA) technology of command register access and remote computer control operation, trigger functions can be  flexibly configured  according to the experimental physical goals. Using the trigger system on CSHINE, we carried out the beam experiment of 25 MeV/u $ ^{86}{\rm Kr}+ ^{124}{\rm Sn}$ on the Radioactive Ion Beam Line 1 in Lanzhou (RIBLL1), China. The online results demonstrate that the trigger system works normally and correctly.  The system can be extended to other experiments.

\end{abstract}


\end{frontmatter}


\section{Introduction}\label{sec. I}
Trigger system is of significance in intermediate and high energy nuclear physics experiment. The reasonable setting of trigger conditions can greatly improve the detection efficiency of physical events of interest and suppress the background events. In intermediate energy heavy ion collisions, the final reaction products are abundant and show $4\pi$ emission in the center of mass system ~\cite{T.Li-1993}, leading to a very rich variety of trigger signals of interest in experiment. Conventionally, the trigger system of the experiment in Fermi energy region is composed of independent analog electronic modules, which have obvious disadvantages of high power consumption, large space occupancy, lack of remote operation and lack of scalability etc. Field programmable gate array (FPGA) is a semi-custom circuit in the field of application specific integrated circuits ~\cite{S.Brown-2007}. It can change the circuit structure. Advantageously, it overcomes the non reconfigurability of application specific integrated circuits and makes up for the limited number of gate circuits of other programmable logic chips ~\cite{E.Monmasson-2007}. It has become a mainstream programmable logic device. The trigger system based on FPGA, is preferable for the advantages of small space, low power consumption, strong adaptability and good scalability, and is convenient for remote control operation ~\cite{V.Lindenstruth-2004}. So the FPGA-based trigger system is gradually adopted by many nuclear physics experiments.

At present, FPGA has been widely used in experimental signal processing as well as  trigger construction, such as the digital signal processing algorithm module ~\cite{Z.Szadkowski-2006, E.Imbergamo-2008, Y.Y.Liu-2018}, time to digital converter (TDC) module ~\cite{M.Bogdan-2005, T.Suwada-2015}, and trigger module in experiment ~\cite{G.Gratta-1997, M.Cambiaghi-2001, F.Clemencio-2019}. For example, FPGA has been successfully used on ATLAS ~\cite{J.Garvey-2003, A.Khomich-2006}, TOTEM ~\cite{M.G.Bagliesi-2010} and CMS ~\cite{M.Jeitler-2013} at LHC, Fermilab SeaQuest experiment ~\cite{S.H.Shiu-2015}, MEG experiment ~\cite{A.Baldini-2016}, LUX dark matter experiment ~\cite{D.S.Akerib-2016}, the external target experiment in the HIRFL-CSR ~\cite{M.Li-2016}, the CLAS12 experiment ~\cite{B.Raydo-2020} and the FTB project ~\cite{P.Ottanelli-2021}. In the above experiments, the electronic module based on FPGA obviously overcomes the disadvantages of the traditional electronics.

This paper describes a trigger system based on FPGA technology, adopting the CAEN V2495 VME programmable logic unit ~\cite{CAEN-V2495} which is a general purpose FPGA and I/O unit housed in a 1-unit wide VME 6U crate. The module performs sequential logic processing on digital signals and digital functions such as coincidence, trigger process, Gate and Delay generator, I/O register etc. Because the electronic module has flexible function expansion and provides a variety of I/O interfaces compatible with other VME electronic modules, it has been applied to the beam experiment at the Compact Spectrometer for Heavy Ion Experiment (CSHINE). The paper is arranged as following. Section \ref{sec. II} presents the structure of CSHINE, the setup of various types of detectors and the relevant front end electronics. Section \ref{sec. III} introduce  the V2495 module briefly and the trigger core logic architecture used in the experiment. Section \ref{sec. IV} presents the performance of the system in the beam experiment, including the timing relationship of the signals involved in the trigger construction. The online experimental results are also presented. Section \ref{sec. V} is the summary.

\section{The setup of CSHINE} \label{sec. II}

\begin{figure}[htb] 
\centering
\includegraphics[width=0.5\textwidth]{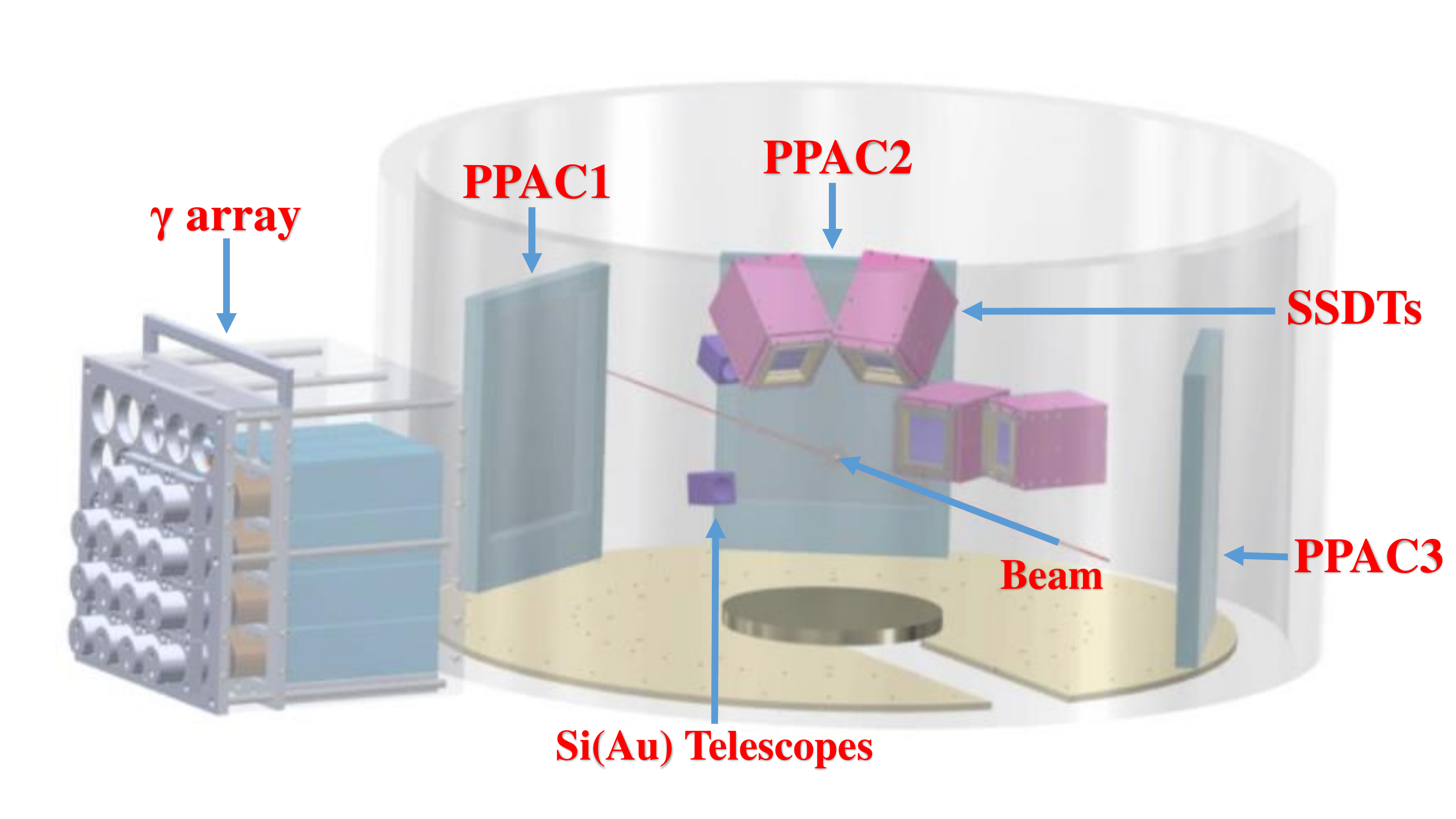} 
\caption{(Color online) Schematic diagram of CSHINE setup for beam experiment in 2022}
\label{CSHINE_2022}
\end{figure}

CSHINE aims at the experimental studies of heavy ion reactions and nuclear equation of state at Fermi energies. It is designed to measure  fast fission fragments (FFs) with the coincident emission of charged particles and high energy $\gamma$-rays. In the current configuration, CSHINE consists of the following detectors: (i) Four sets of silicon strip detector telescopes (SSDTs) to measure the light charged particles (LCPs) and the intermediate mass fragments (IMFs), (ii) Three Parallel Plate Avalanche Counters (PPACs) to measure the fission fragments, (iii) One electromagnetic calorimeter of CsI crystal ($\gamma$ hodoscope) to record the high-energy $\gamma$ rays produced via bremsstrahlung  at the early stage of collisions, and iv) 2 Si(Au) telescope at small angle to record the elastic scattering of projectile. The setup of CSHINE  is shown in Fig.\ref{CSHINE_2022}. Each sub-detector is introduced below.

\subsection{CSHINE  Sub-detectors}\label{sec. II1}

The LCPs and IMFs in coincidence with the fission fragments are measured by the SSDTs, each of which is composed by a thin single-sided silicon strip detector (SSSSD) and a thick double-sided silicon strip detector (DSSSD) backed by a 3 $\times$ 3 CsI(Tl) crystal array, delivering the energy loss $\Delta E_1$, $\Delta E_2$ and deposit energy $E_{\rm CsI}$, respectively. At the entrance, a 2$\mu m$ incident window of aluminium coated Mylar film is mounted  to stop the delta electrons. The detailed structure of the SSDTs can be found in ~\cite{F.H.Guan-2021}. An adapter made of PCB plate is mounted at the end of the telescope to connect and transfer the signals of SSSSD, DSSSD and CsI array to the front-end electronics, respectively. The timing signal (T signal) of the incident particle is extracted from the DSSSD layer. The LCPs and IMFs are identified by $\Delta E-E$ method ~\cite{F.H.Guan-2022}.

 PPACs are mounted to measure the timing and position information of FFs emitted in heavy ion reactions. It is a two-dimensional position sensitive gas detector composed of a cathode layer  in the middle and mutually perpendicular wire anode layers on both sides. The working gas is $\rm C_{4} \rm H_{10}$ ~\cite{F.H.Guan-2021}. The negative high voltage is applied to the cathode plane, which delivers the timing signal. The signals induced on the anode planes are transferred to both ends by delay line, and the time differences are used to deduce the incident position  for both $x$ and $y$ dimensions, respectively. The first PPAC (PPAC1) is the main fission fragment detector,  while the second (PPAC2) and the third (PPAC3) ones are installed at the front and middle angles on the other side of the beam as coincident fragment detectors.

The $\gamma$ hodoscope  consists of 25 CsI(Tl) scintillators in 5 $\times$ 5 configuration (in the current experiment, only 15 units are installed). The size of each CsI(Tl) scintillator is 70 $\times$ 70 $\times$ 250 $mm^{3}$. The optical signal generated in the crystal is collected and amplified by the photomultiplier tubes (PMTs). The high voltage applied to the PMTs ranges  from 750V to 1kV. 

\subsection{CSHINE Physical Goals}\label{sec. II2}

The physical goals of CSHINE include, but are not limited to the following studies related to the the equation of state of asymmetric nuclear matter,  i.e., the density dependence of the nuclear symmetry energy ~\esym, which is of significance in both nuclear physics and astrophysics  ~\cite{M.Colonna-2020, B.A.Li-2014}: (i) The isospin effect of the emission of charged particles in heavy ion reactions. It is shown, for instance, the angular distribution of neutron-richness of LCPs reflects the isospin-dependent emission hierarchy and probes the stiffness of ~\esym~ \cite{R.S.Wang-2014, Y.Zhang-2017, B.A.Li-2018, J.Xu-2019, B.A.Li-2021}. And hence, isospin chronology based on correlation function measurement of particle pairs is required to determine quantitatively the emission sequence  of particles of different $N/Z$ composition, from which one can infer the transport of isospin degree of freedom ~\cite{G.Verde-2006, Y.J.Wang-2021, Y.J.Wang-2022} arising from the effect of \esym ~\cite {Z.G.Xiao-2006}. (ii) Studies on the dynamic feature of the fission process. The properties of the fast fission induced by heavy ion reactions can be studied by measuring the fission fragments and the coincident LCPs ~\cite{X.Y.Diao-2022}, such that the nuclear symmetry energy can be inferred because of the formation and persistence of the low-density and neutron-rich neck, according to transport model simulations ~\cite{Q.H.Wu-2020, Q.H.Wu-2019}. (iii) The high energy $\gamma$ rays emitted in the early stage of heavy ion  collision. As predicted by some theoretical simulations, the intensity of the high energy $\gamma$ rays shows significant dependence on \esym ~\cite{G.C.Yong-2008, Z.G.Xiao-2014}.

Therefore, CSHINE needs to obtain the following types of the event: (1) The two-body LCPs from the SSDTs for the correlation function measurement, (2) The coincidence  of two heavy fission fragments in PPACs, defining the fission events, (3) The coincidence between fission event and one LCP from SSDTs, (4) The coincidence between fission event and one $\gamma$-ray emission and (5) The correlation between one-body $\gamma$  and one-body LCP.  In addition, inclusive events from all detectors shall be triggered for detector calibration. The above requirements define the trigger scheme of CSHINE.

\begin{figure*}[t] 
\centering
\includegraphics[width=0.78\textwidth]{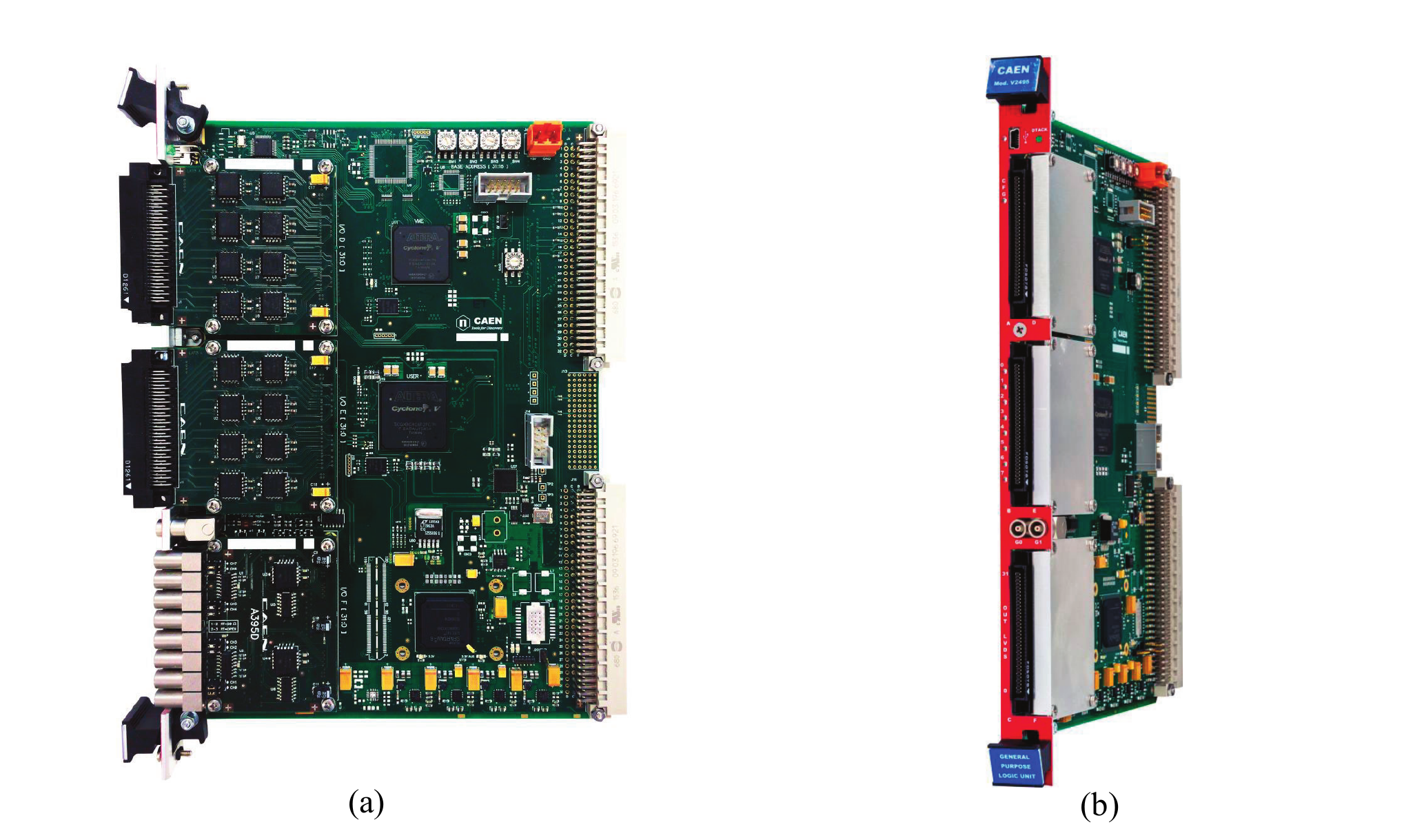} 
\caption{(Color online) (a) real photo of the side view of V2495 module and (b) is real photo of the front view of V2495 ~\cite{CAEN-V2495}}
\label{real_V2495}
\end{figure*}

\section{System Architecture} \label{sec. III}
\subsection{Front-end module}\label{sec. III1}

For the SSDTs, the signals from all channels of the SSSSD, the DSSSD and the CsI units are all transferred from the end adapter plate to Mesytec MPR-16 pre-amplifiers and then successively to Mesytec MSCF-16 main-amplifiers. The timing and energy output signals of the main amplifiers are transferred to TDC (V775/N, CAEN) and ADC (V785, CAEN) respectively. At the front panel of each MSCF-16 corresponding to the DSSSDs, there is one NIM connector delivering a trigger output signal (T-out) if any of the 16 channels is fired. This T-out signal is used as a one-body LCP indicator. Besides, the multiplicity output (M-out ) via a NIM connector on the rear panel delivers a linear signal representing the multiplicity of the 16 input channels. The amplitude of the M-out is $M\times 100$ mV , where $M$ is the multiplicity. The M-out signal from all DSSSD amplifiers are summed further in a linear fan-in and fan-out module to generate an $M_{\rm tot}$ signal containing the information the total multiplicity of LCP from all SSDTs. The CF8000 module to receive this $M_{\rm tot}$ signal can deliver a trigger for two-body LCP events with a threshold setting above 150 mV ~\cite{F.H.Guan-2021}.  

PPAC front-end electronics deals with the timing signal T from cathode and position signals $X_{1}$, $X_{2}$, $Y_{1}$ and $Y_{2}$ from anodes of the PPAC detector. The original signals of PPAC are first amplified by ORTEC Fasting Timing Amplifier (FTA 820) and then input to CF8000 module to generate the NIM logic signals. The output signals of the CF8000 are delayed by GG8020 (Gate Generator) before they are digitized by TDC (V775/N, CAEN). Besides, the timing signal output from CF8000 of each PPAC is split to the second path to construct the trigger signal for fission events. Two types of coincidence are constructed according to the location of the three PPACs, the coincidence of PPAC1 and PPAC2 (marked as PPAC1 $\times$ 2) and that of PPAC1 and PPAC3 (marked as PPAC1 $\times$ 3), respectively ~\cite{F.H.Guan-2021,X.Y.Diao-2022}. Since PPAC2 and PPAC3 are located on the same side of the beam,  so the coincidence between PPAC2 and PPAC3 are not set.

The $\gamma$ array is read out by photomultiplier tubes (PMTs). The anode output of each PMT enters the preamplifier CAEN N914, which delivers an energy output and a fast timing output. While the energy output are fed to the ADC, all the fast time signals of the 15 units are fed into V2495 for a twofold purpose. One is to generate one-body $\gamma$ event, the other is to fan out a signal in LVDS format, which is converted to ECL and transferred to TDC (V775/N, CAEN) for data acquisition.

\subsection{V2495 module}\label{sec. III2}

The V2495 module is produced by CAEN. It is mainly composed of three FPGA chips, one I/O unit and a 50 MHz clock crystal oscillator module. It is installed in a one-unit wide VME 6U standard module. Fig.\ref{real_V2495} presents a real photo of the side view (a) and a front view (b) of a V2495 module, respectively ~\cite{CAEN-V2495}.

\begin{figure}[H] 
\centering
\includegraphics[width=0.45\textwidth]{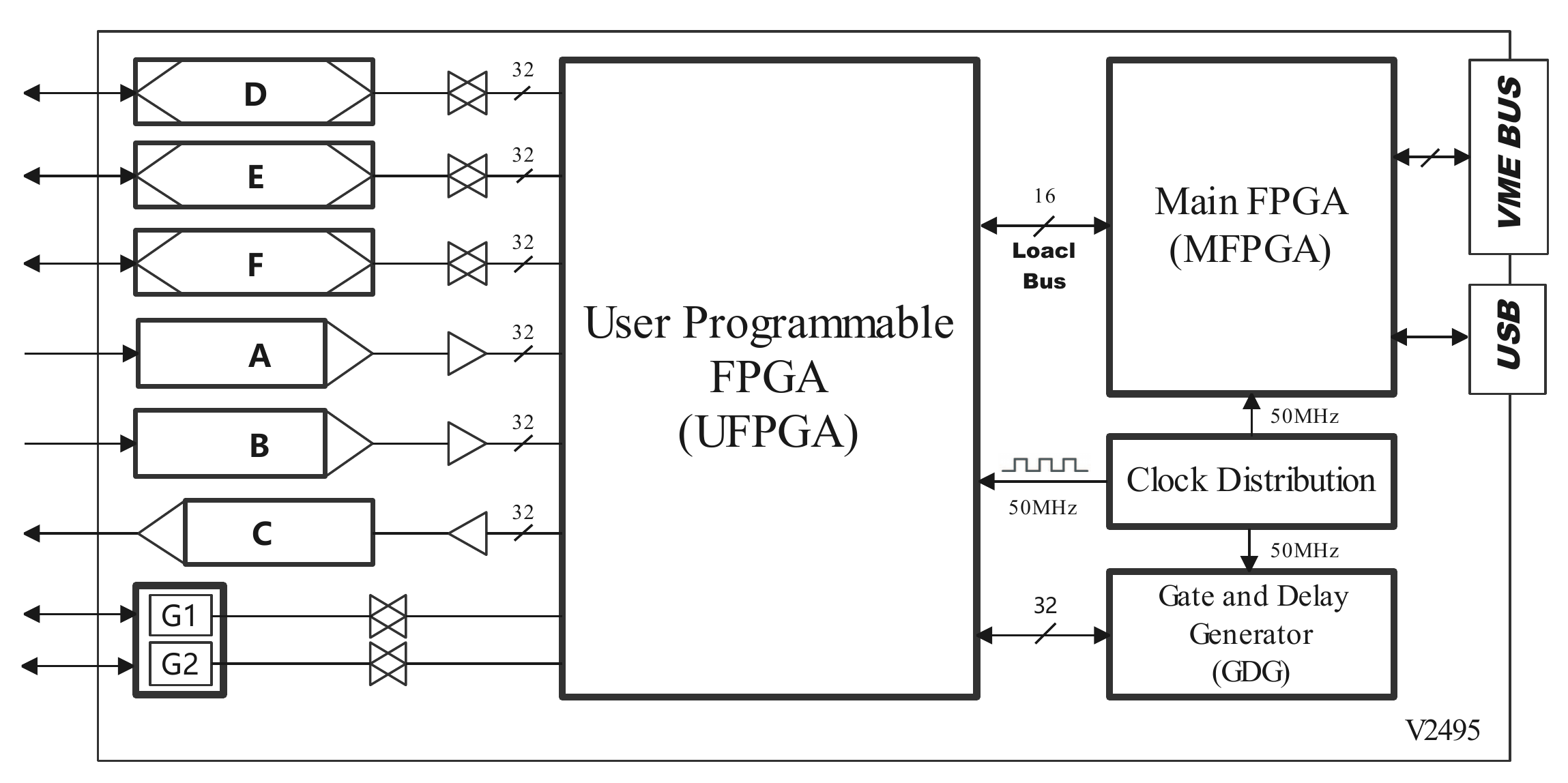} 
\caption{(Color online) The architecture of the FPGA connections of V2495}
\label{architecture_V2495}
\end{figure}

The FPGA architecture of V2495 is shown in Fig.\ref{architecture_V2495}. The 3 FPGA chips are listed as following. The main FPGA (MFPGA) is mainly used for external communication of V2495, for which VME bus protocol and USB protocol can be used. The user FPGA (UFPGA) chip model is Altera cyclone V 5CGXC4, which contains 50K logic units, 162 input interfaces and 130 output interfaces. Xilinx spartan-6 FPGA chip realizes  burning the gate and delay generation (GDG) module. The expansion slot of I/O module connected to the physical pin of UFPGA chip in the module enables V2495 to support the input and output of LVDS, ECL, NIM and TTL signals by adding three independent sandwich boards ~\cite{CAEN-V2495}. 

\begin{figure}[H] 
\centering
\includegraphics[width=0.5\textwidth]{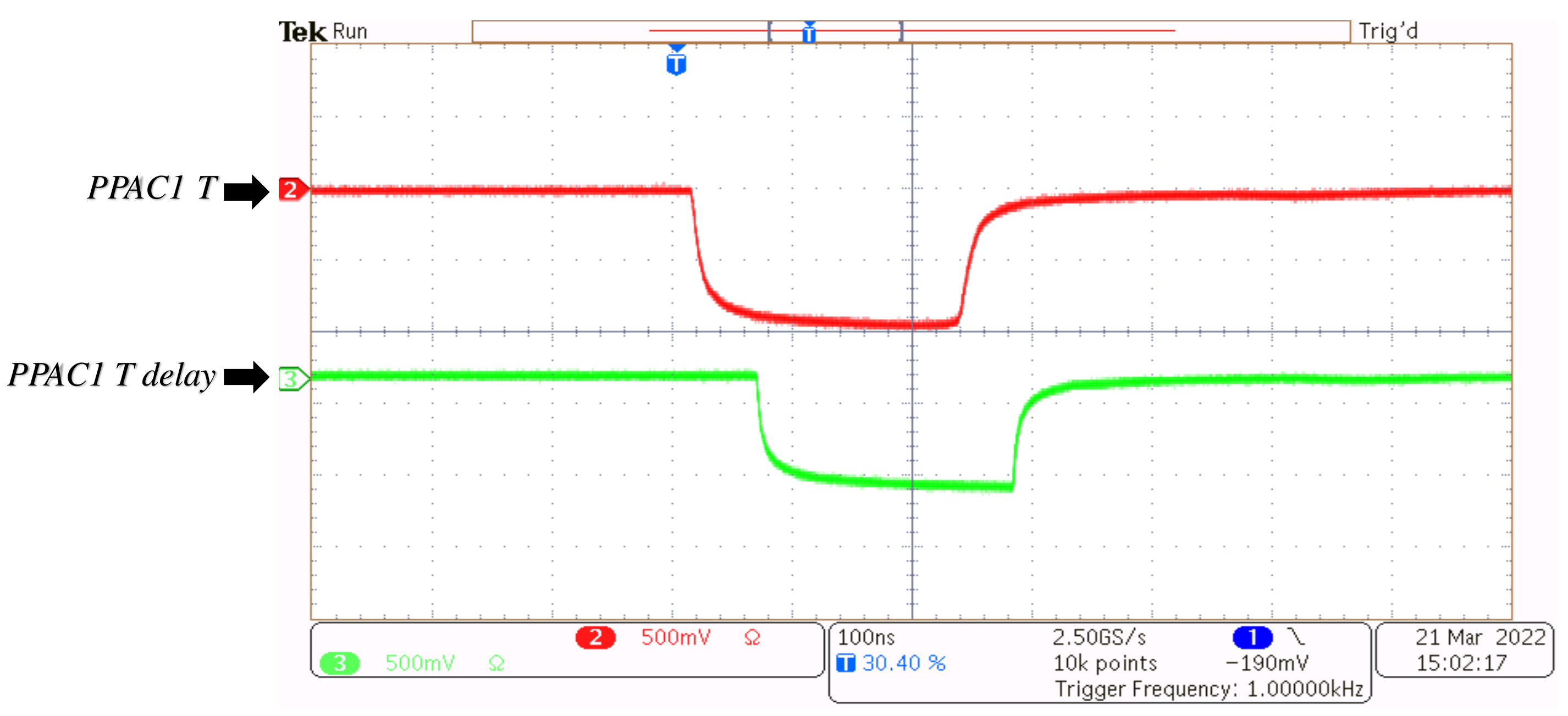} 
\caption{(Color online) Sample oscilloscope diagram of PPAC1 T signal delay, in which $PPAC1 T$ represents the original signal and $PPAC1 T delay$ represents the output signal after passing through the GDG module}
\label{PPACT_signal_delay}
\end{figure}

The GDG module is used to control signal gating, delaying and broadening. The module can provide 32 signal interfaces. In the preparation stage of the beam experiment of CSHINE, Specific signals generated by pulser are fed to the front-end electronics of each detector system to simulate the experimental output, in order to conduct timing calibration and to test the functions of the trigger module. As an example, Fig.\ref{PPACT_signal_delay} shows the delay of T signal of PPAC1 in pulser test. The original signal enters the GDG module and is delayed by  $40~ns$ through remote control.

\begin{figure*}[b] 
\centering
\includegraphics[width=1.0\textwidth]{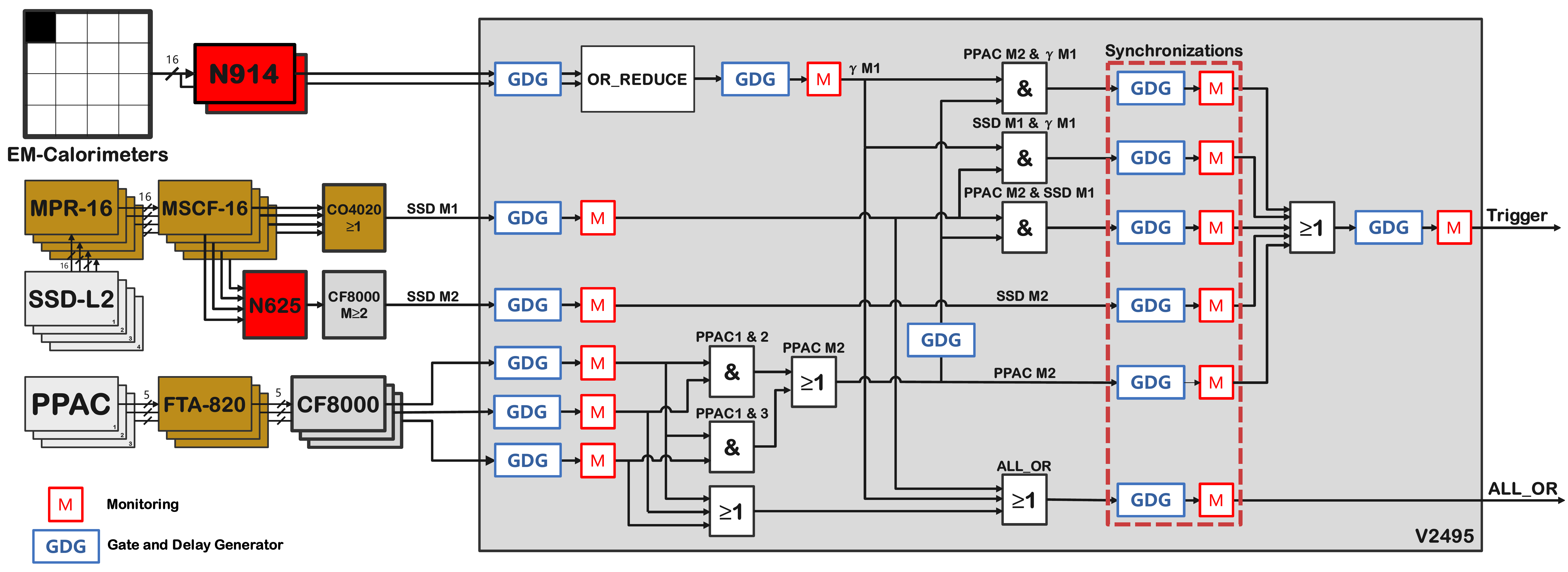}
\caption{(Color online) The logic schematic diagram of CSHINE trigger system}
\label{CSHINE_trigger_logic}
\end{figure*}

\subsection{Trigger signal processing}\label{sec. III2}

\subsubsection{Trigger Logic Unit}\label{sec. III2}

The logic processing in the  trigger system of CSHINE is compiled in Quartus II 13.0 by VHDL, and the remote logic update of the trigger system is realized in UFPGA by CAEN electronic firmware upgrade tool CAENUpgrader~\cite{CAEN-V2495,CAEN-VX2495-User}.

Fig.\ref{CSHINE_trigger_logic} presents the logic schematic diagram of the trigger system. The output of the front-end electronics of sub-detectors are used to produce the logic signals before being fed to the trigger module V2495, where the trigger scheme is constructed with the following operations: (1) The fast time signal of PMTs from N914 is used to generate one-body $\gamma$ signal ($\gamma ~M1$). (2) Fission event signal (PPAC $M2$) is defined by the OR calculation of ${\rm PPAC 1\times 2 }$ and ${\rm PPAC 1\times 3 }$, corresponding to the coincidence between the main FF detector PPAC1 and the second FF detector,  PPAC2 and PPAC 3, respectively.  (3) The two-body LCP signal is given by SSD $M2$ signal. (4) Total one-body trigger signal (ALL OR) for test and detector calibration is obtained inclusively by OR operation of one-body $\gamma$ signal, one-body FF signal from PPAC  (PPAC $M1$) and one LCP signal ( SSDT $M1$).  Finally,  the global trigger signal of CSHINE experiment covers the event types listed below, which are generated by selections or coincidences of the above signals synchronized by applying proper delay to each. 

- Fission event:  PPAC  $M2$

- Fission \& 1 LCP: PPAC $M2$ .and. SSDT $M1$ 

- Fission \& 1 $\gamma$: PPAC $M2$ .and. $\gamma ~M1$

- 2-body LCP coincidence:   SSDT $M2$

- LCP \& $\gamma$ coincidence: SSDT $M1$  .and. $\gamma ~M1$

- Inclusive (All OR):   SSDT $ M1$ .or. $\gamma ~M1$ .or.  PPAC $M1$

\subsubsection{Timing relationship and Simulation of signal}\label{sec. III2}

\begin{figure}[t] 
\centering
\includegraphics[width=0.51\textwidth]{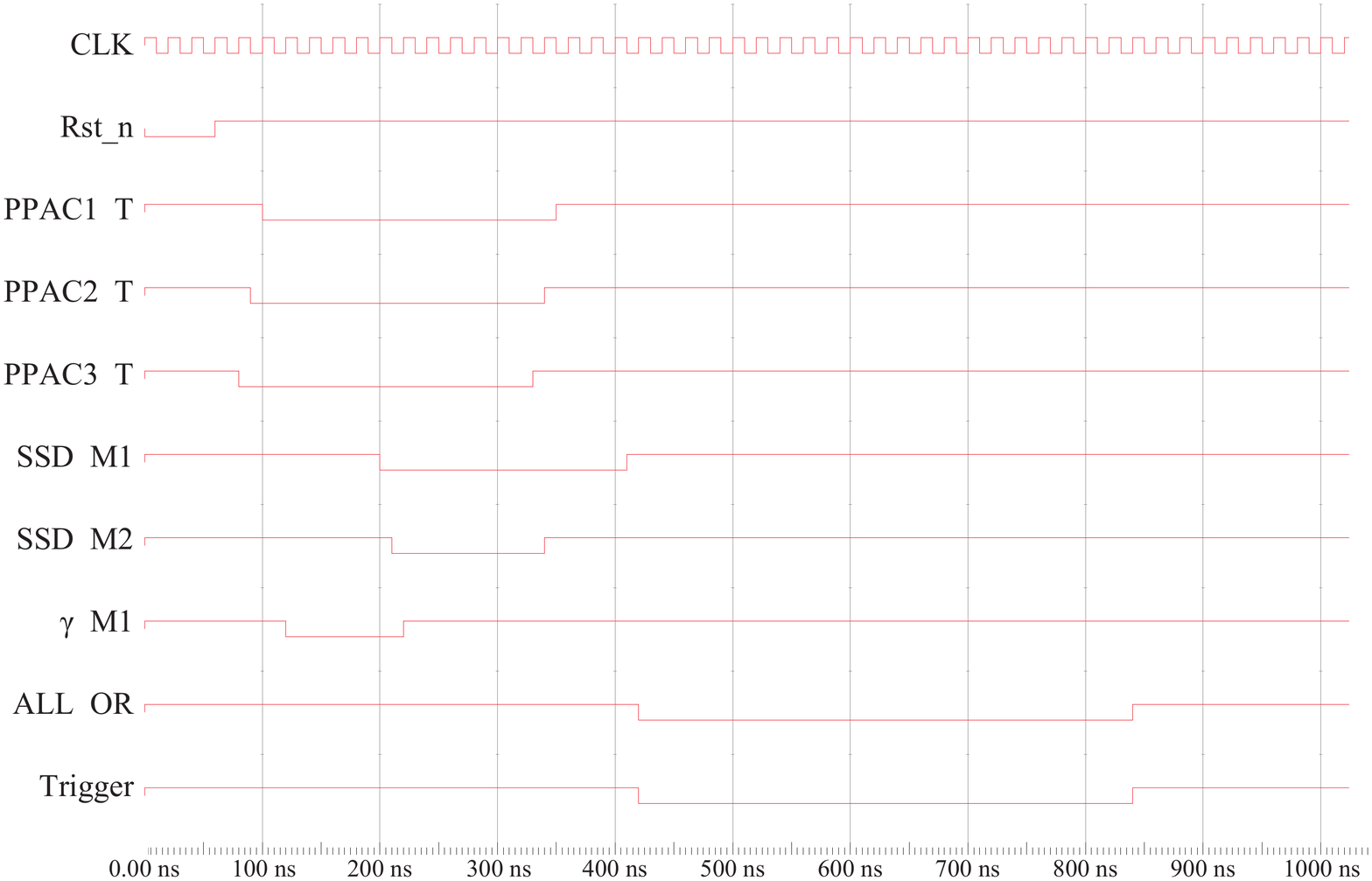} 
\caption{(Color online) Simulation waveform diagram of trigger signal timing relationship}
\label{Sim_CSHINE_trigger}
\end{figure}

In order to coincide the fission event  with other types of signals and synchronize various output trigger signals, it is necessary to set proper delay and broadening to all types of trigger via the GDG module, as depicted by GDG boxes in Fig.\ref{CSHINE_trigger_logic}. Before beam experiment, in order to obtain the correct time sequences of the trigger signal, we conducted the simulations for the whole trigger logic scheme in the framework of Quartus II 13.0 and ModelSim 13.0.  In the simulation, the parameters of the delay time and  broadening length of the input signals participating in the trigger construction are optimized by considering the response delay of different types of detectors and the front end electronics. For instance, the SSDT $M1$ signal is later than the PPAC timing because the former is extracted from the Mesytec MSCF-16. As an example, the parameters of the delay and width  for the experiment ${\rm ^{86}Kr+^{208}Pb}$ at 25 MeV/u are listed  in Table \ref{Delay and widen para} and Table \ref{Trigger signal para}. Fig. \ref{Sim_CSHINE_trigger} presents the time sequences of the trigger signals with the settings. The PPAC timing signal, the SSDT $M1$ and $M2$ and the $\gamma$ timing signal are generated as input signals, where the relative delays to each other are chosen according to the real response of various detectors.  With the parameter settings, it is shown that the trigger signal are correctly obtained, holding 300 ns delay to the arrival of PPAC timing signal. 

\begin{table*}[htb]
\caption{Delay and width of each input signal and total one-body trigger signal.}
\label{Delay and widen para}
\setlength\tabcolsep{13pt}
\centering
\small
\begin{tabular}{lcccccccc}
\toprule
signal type & PPAC1 T   & PPAC2 T  & PPAC3 T  & PPAC M2  & SSD M1  & SSD M2  & $\gamma$ M1  & ALL OR \\[1.5pt] 
\midrule
delay($ns$)  & 20 & 20 & 20 & 0 & 20 & 20 & 20 & 200 \\[1.5pt]
width($ns$)  & 400 & 400 & 400 & 640 & 80 & 320 & 80 & 400   \\[1.5pt]
\bottomrule
\end{tabular}
\end{table*}

\begin{table*}[htb]
\caption{ Delay and broadening of the trigger signals corresponding to difference events types.}
\label{Trigger signal para}
\setlength\tabcolsep{13pt}
\centering
\small
\begin{tabular}{lcccccc}
\toprule
signal type  & PPAC M2  & SSD M2  & PPAC M2 \& SSD M1   & PPAC M2 \& $\gamma$ M1  & SSD M1 \& $\gamma$ M1  & Trigger \\[1.5pt] 
\midrule
delay($ns$)  & 200 & 200 & 320 & 0 & 0 & 0 \\[1.5pt]
width($ns$)  & 200 & 320 & 320 & 320 & 320 & 400 \\[1.5pt]
\bottomrule
\end{tabular}
\end{table*}

\subsection{Control and firmware update}\label{sec. III3}

The remote control and logic function update of the trigger system adopts the CAEN Chain Optical Network protocol (CONET). The upper computer transmits signals and instructions to V2718 in slot 0 of the VME crate through PCI/PCIe protocol and CONET protocol. The V2718 interacts with V2495 through VME backplane bus ~\cite{PLU-Lib}. The upper computer operating system used in this experiment is scientific Linux 6.4 i386.

Working with Linux system on the upper computer, Quartus II 13.0 is installed to compile the files of the trigger system and generate  roleplay designer (RPD) files, which are then burnt into the  UFPGA of  V2495  through VME bus. The communication of the command and data flow between the host computer to the VME crate is implemented  by the optical fiber. Since the VME crate housing the digitization modules are usually located on the experimental site, therefore, the configuration and functioning of the trigger system is indeed  remotely  controlled. The remote configuration program, designed in C++ and run on upper computer,  calls CAEN PLU library and configures the GDG module of V2495 through SPI interface ~\cite{PLU-Lib}. 

\subsection{Data acquisition system}\label{sec. III4}

This experiment uses the VME standard data acquisition (DAQ) system. The system is operated based on the following hardware: the ADC (V785, CAEN), TDC (V775/N, CAEN), and CAEN A2818 communication PCI board. The DAQ software is compiled based on C++ on the upper computer, relying on CAENDigitizer Library and CAENComm library to complete the driver configuration of the DAQ system. The online display of the experimental spectra is realized by embedding ROOT library. 

\begin{figure}[htb] 
\centering
\includegraphics[width=0.4\textwidth]{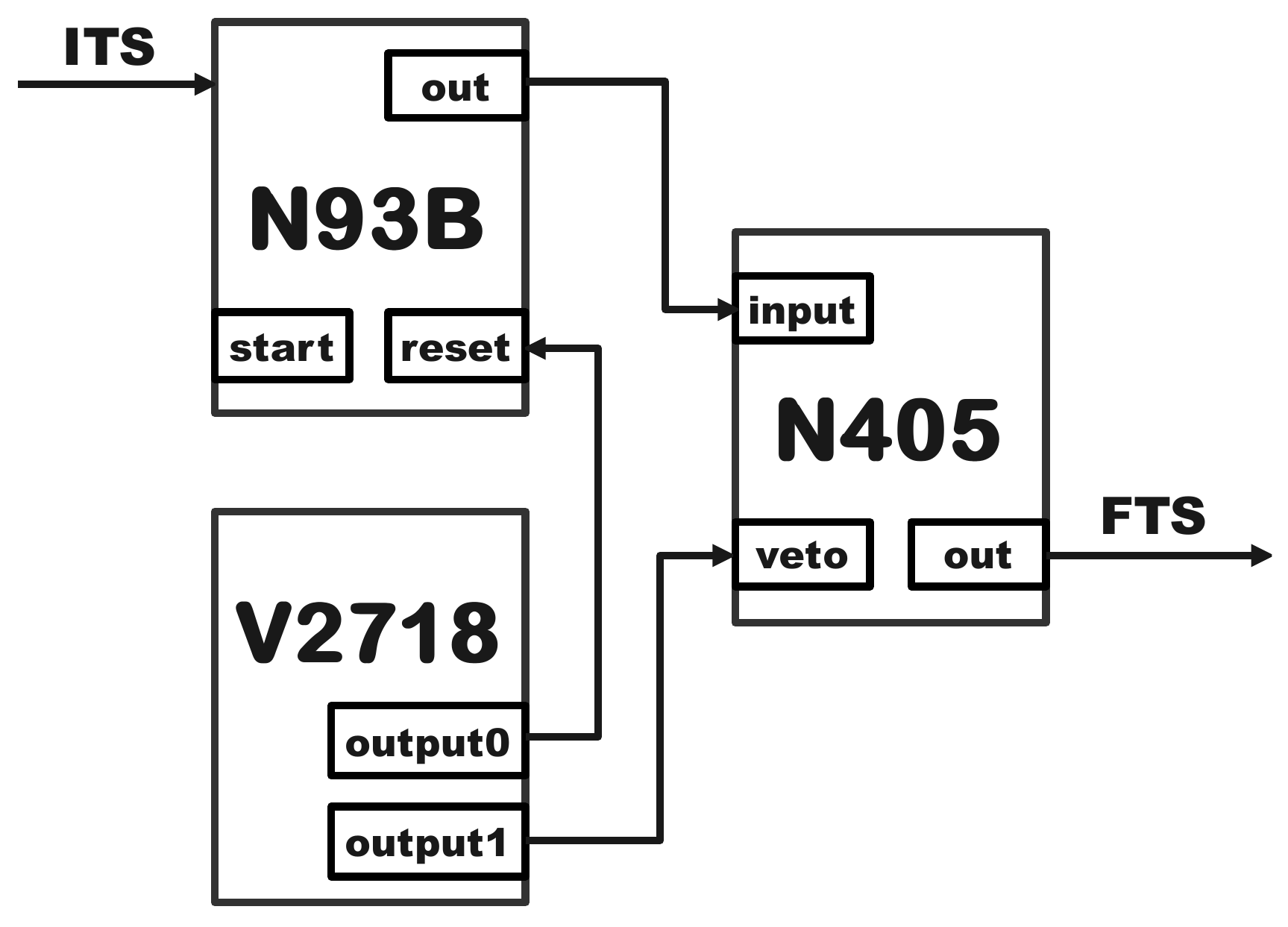} 
\caption{DAQ system interrupt circuit. ITS is the initial trigger signal and FTS is the final trigger signal.}
\label{interrupt_circuit}
\end{figure}

The DAQ system responds to the trigger signal real-time before starting signal digitization and data transfer. But if in any case, a next trigger signal arrives before the previous one is completely processed, data acquisition error occurs. Therefore, it is the user's responsibility to set up an external interrupt circuit for DAQ system to avoid such conflict.  CAEN N93B and N405 electronic modules are used to build the external interrupt circuit. As shown in Fig.\ref{interrupt_circuit}, the time switcher of the N93B module is set to $\infty$, i.e., the output signal of N93B is always at logic 1 unless a {\it reset} signal is received. Such {\it reset} signal is sent from the {\it Output}-0 port of V2718  when the digitized data are all save to the storage.   When the system is in busy status, however,  {\it Output}-1 at NIM high level is sent from the V2718 to the {\it veto} port of N405. This ensures that during the dead time of DAQ system, even if a next  experimental trigger (initial) signal arrives, the external interrupt circuit conducts an anti coincidence operation and blocks the output of the final trigger for the later event from the  N405.Therefore, V2718 provides the DAQ system with the busy signal corresponding to the dead time.

\begin{figure}[!h] 
\centering
\includegraphics[width=0.45\textwidth]{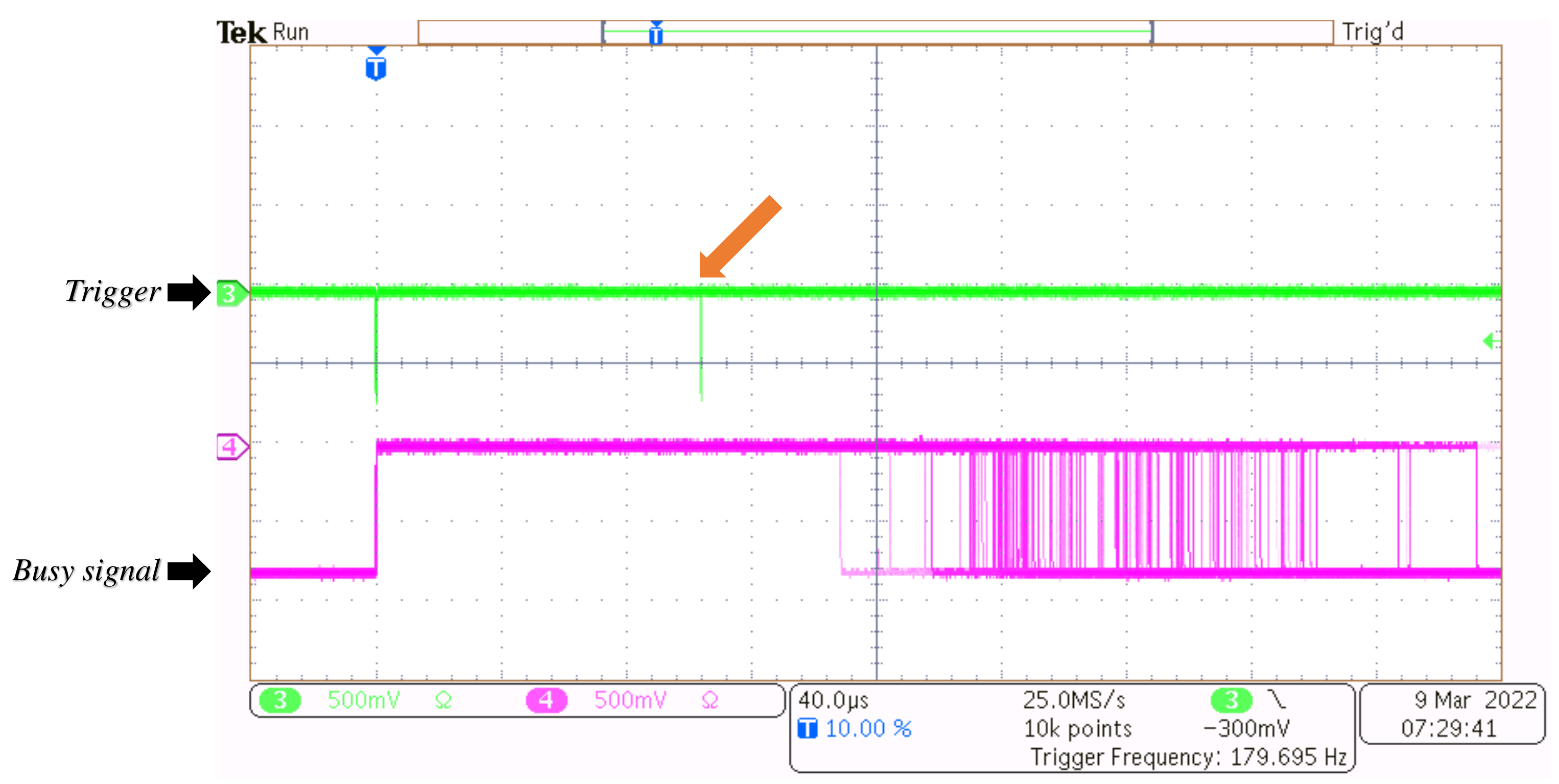} 
\caption{(Color online) Trigger signal and DAQ busy signal during beam experiment. The display is in persistence mode. }
\label{dead time}
\end{figure}

Fig.\ref{dead time} is the oscilloscope display of  the trigger signal (upper, green) and DAQ busy signal (lower, purple) detected during the beam experiment. We note that the display is in the persistence mode.  As can be seen from the figure, once the first trigger signal arrives, DAQ busy signal is enabled. During the dead time period, if a next trigger signal arrives, the system does not respond. As indicated by the arrow in the figure, the second trigger signal is shielded since it arrives during the dead time.  It is shown that, for each triggered event, the total processing time of DAQ is about 200 $\mu$s, depending on the number of firing channels.


\section{Beam experiment results} \label{sec. IV}

\begin{figure}[H] 
\centering
\includegraphics[width=0.45\textwidth]{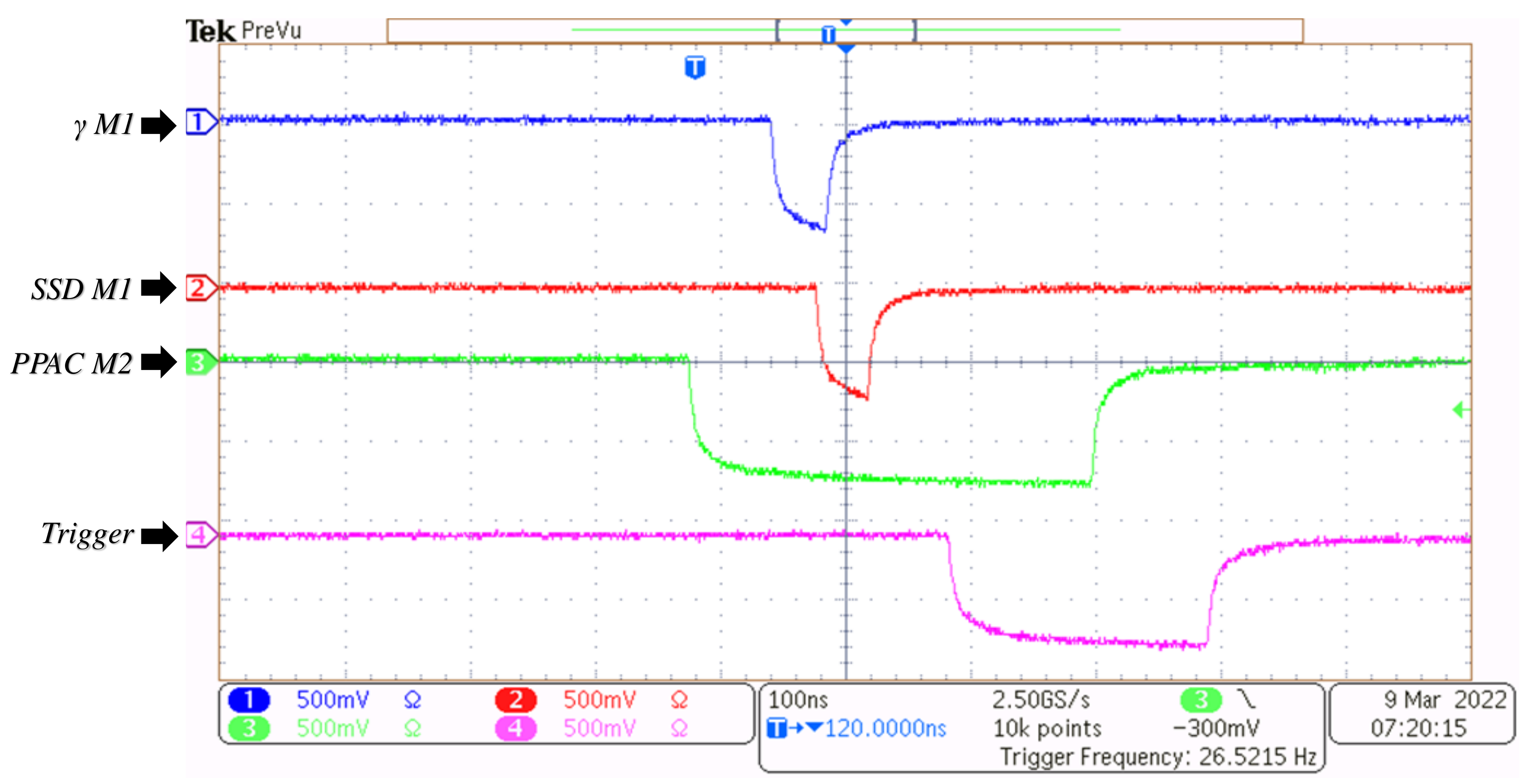} 
\caption{(Color online) Timing relationship between the input signals and the final trigger signal during beam experiment.}
\label{trigger}
\end{figure}

During the beam experiment, the timing relationship and time width of various input and output signals are configured in the trigger system. The fission event is among  the trigger conditions, because the geometry of the two-body fission events define the event topology of physical interest. Fig.\ref{trigger} shows a typical event with the timing relationship between each detector signal and the total trigger signal obtained during the beam experiment. In this event, the PPAC $M2$ corresponding to fission event (green), $\gamma ~M1$ (blue) and SSDT $M1$ (red)  are all presented. As a result, the trigger signal (purple) is correctly generated, delayed by 200 ns  from the generation of PPAC $M2$ signal, which is consistent with the signal simulation results. It is demonstrated  that the trigger circuit works correctly as expected. In this experiment, the total one-body counting rate is in range of ${\rm 20-40~k ~(s^{-1})}$, and the global trigger rate is in range of ${\rm 0.5-1~k ~ (s^{-1})}$ depending on the beam intensity. The DAQ dead time is in range of ${\rm 80-200~\mu s}$.

\begin{figure}[htb] 
\centering
\includegraphics[width=0.45\textwidth]{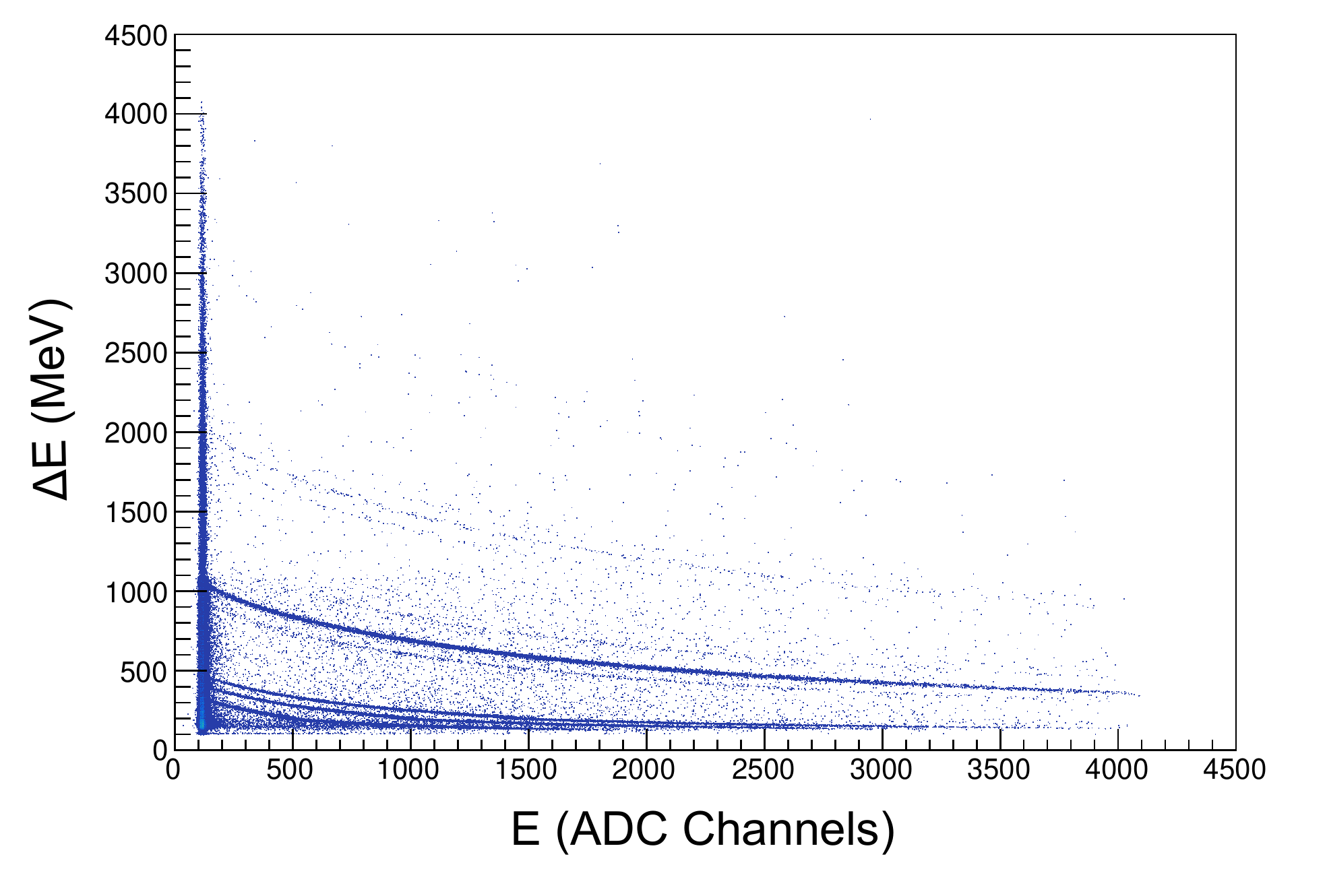} 
\caption{(Color online) Online display of the $\Delta E_2-E_{\rm CsI}$ scattering plot of SSDT 4}
\label{DE_E}
\end{figure}

Fig.\ref{DE_E} shows an online $\Delta E_2-E_{\rm CsI}$ scattering plot of SSDT4, where $\Delta E_2$ is from the DSSSD and $E_{\rm CsI}$ is from the CsI crystal. It can be seen that all the isotopes of $Z\le 3$ are clearly identified on the online plot. The preliminary analysis results of the experimental data show that the trigger system works normally in CSHINE and meets the experimental requirements.

\section{Conclusion} \label{sec. V}

An expandable trigger system that meets the physical requirements has been designed for CSHINE. The system adopts V2495 logic unit and realizes the trigger functions in FPGA equipment. By designing  and modifying trigger functions according to user commands or online reconfiguration of FPGA, the signal timing sequences meet the experimental requirements and good flexibility of the system can be guaranteed. The functionality of the system is demonstrated in the beam experiment with CSHINE  on the RIBLL1 in 2022. This work provides a practical trigger solution for CSHINE and other nuclear physics experiments in similar scale.

\section*{Declaration of competing interest}
The authors declare that they have no known competing finan-cial interests or personal relationships that could have appeared to influence the work reported in this paper.

\section*{Acknowledgments}
This work is supported by the National Natural Science Foundation of China under Grants Nos. 11875174, 11961131010 and 11961141004, by the  Initiative Scientific Research Program of Tsinghua University, and by the Heavy Ion Research Facility in Lanzhou (HIRFL). The authors acknowledge the RIBLL group for offering local help in experiment and the machine staff for delivering the beam. 

\end{document}